\documentclass[twocolumn]{aas}

\usepackage{bm}         % for bold math symbols
\usepackage{amsmath}    % for equation* environment
\usepackage{CJKutf8}    % for Chinese characters
\usepackage{silence}    % to filter out warnings

\defcitealias{bi.J.2021.05}{paper I}

\begin{document}

\title{Puffed-up Edges of Planet-opened Gaps in Protoplanetary Disks. II. \\ The Role of the Planet's Orbital Eccentricity}

\author[orcid=0000-0002-0605-4961,sname='Bi',gname='Jiaqing']{Jiaqing Bi \begin{CJK*}{UTF8}{gbsn}(毕嘉擎)\end{CJK*}}
\affiliation{Max-Planck-Institut für Astronomie, 69117 Heidelberg, Germany}
\affiliation{Academia Sinica Institute of Astronomy and Astrophysics, Taipei 106319, Taiwan, Republic of China}
\email[show]{bi@mpia.de}

\author[orcid=0000-0002-8597-4386,sname='Lin',gname='Min-Kai']{Min-Kai Lin \begin{CJK*}{UTF8}{bsmi}(林明楷)\end{CJK*}}
\affiliation{Academia Sinica Institute of Astronomy and Astrophysics, Taipei 106319, Taiwan, Republic of China}
\affiliation{Physics Division, National Center for Theoretical Sciences, Taipei 106319, Taiwan, Republic of China}
\affiliation{Tsung-Dao Lee Institute, Shanghai Jiao Tong University, Shanghai 201210, People's Republic of China}
\email[show]{mklin@asiaa.sinica.edu.tw}

\begin{abstract}
 
Eccentric planets constitute a large population of known exoplanets and may drive significant substructures in protoplanetary disks through planet-disk interactions if their eccentricities are excited early in the planet formation process. In this paper, we investigate the impact of a planet's orbital eccentricity on gas and dust structures in protoplanetary disks using three-dimensional multifluid hydrodynamic simulations. We find that an eccentric planet can drive stronger meridional gas circulation around the planet-opened gap, which significantly enhances the dust puff-up feature at the gap edge relative to the circular-orbit case. The planet-induced gap can also become highly leaky to dust grains when the planet is eccentric, allowing dust grains to be transported radially and thereby fill the gap. Furthermore, dust rings composed of pebble-sized grains are expected to become both larger and radially wider when the planet is eccentric, with this trend becoming more pronounced at higher planet eccentricities. Overall, our results suggest that a planet's orbital eccentricity can play a significant role in shaping gas and dust structures in protoplanetary disks, with important implications for planet formation theory and disk observations of the WISPIT 2 system.

\end{abstract}


\section{Introduction} \label{sec:intro}

Among the more than 6,000 exoplanets discovered to date\footnote{Exoplanet Archive; https://exoplanetarchive.ipac.caltech.edu/}, giant planets beyond the ``hot'' regime are typically found to exhibit a moderate level of orbital eccentricity. This trend appears to be robust across different detection methods, including radial velocity (\citealt{kane.SR.2024.02,rosenthal.LJ.2024.01,stevenson.AT.2025.05,blunt.S.2026.03}),
% kane.SR.2024.02               % median e = 0.23 for cold-giants beyond snowline
% rosenthal.LJ.2024.01          % median e = 0.23 for multi-giant systems beyond 0.3 au 
% stevenson.AT.2025.05          % mean e = 0.245 for RV planets more than Saturn-mass
% blunt.S.2026.03               % e distribution peaking at 0.3 for super-jovian planets
transiting (\citealt{alqasim.A.2025.05}), 
% alqasim.A.2025.05             % 40% transiting long-period giant planets have e > 0.2
and direct imaging (\citealt{bowler.BP.2020.02,nagpal.V.2023.02,doo.CR.2023.08}).
% bowler.BP.2020.02             % e distribution peaking at 0.13 for super-jovian planets
% nagpal.V.2023.02              % some updates on top of bowler.BP.2020.02
% doo.CR.2023.08                % some updates on top of bowler.BP.2020.02
The prevalence of non-zero eccentricity for those warm and cold giants is not too surprising, as the tidal circularization effect becomes much less efficient at larger orbital radii (\citealt{ogilvie.GI.2014.08}).
% ogilvie.GI.2014.08            % annual review on tidal dissipation
But the mechanisms responsible for exciting the eccentricity in the first place remain under debate.

The eccentric Kozai-Lidov mechanism (EKL; \citealt{naoz.S.2016.09}, and references therein) provides a promising pathway for eccentricity excitation in systems with certain hierarchical configurations. 
% naoz.S.2016.09                % annual review on eccentric Kozai-Lidov mechanism
However, recent statistical analyses of transiting, long-period giant planets (TLGs) suggest that planet-planet scattering may play a more dominant role than EKL. This is because no strong correlation is found between eccentricity and the presence of stellar companions among those TLGs, but an anti-correlation between eccentricity and the planet multiplicity in the system is observed (\citealt{alqasim.A.2025.05}). 
% alqasim.A.2025.05             % statistical analysis on 92 TLGs
While scattering is efficient at producing high eccentricity values, it tends to underproduce planets in the low to moderate eccentricity range (\citealt{rasio.FA.1996.11,papaloizou.JCB.2001.07,chatterjee.S.2008.10,juric.M.2008.10}).
% rasio.FA.1996.11              % early works on planet-planet scattering
% papaloizou.JCB.2001.07        % early works on planet-planet scattering
% chatterjee.S.2008.10          % 1D N-body model of three planets, underproduce the e < 0.2 population
% juric.M.2008.10               % similar conclusions to the one above
In contrast, planet-disk interactions provide a complementary pathway, because they preferentially excite eccentricity values in this range for giant planets (\citealt{dangelo.G.2006.12,duffell.PC.2015.10,ragusa.E.2018.03}). 
% dangelo.G.2006.12             % e grows to 0.11 in 3e4 orbits for 2M_J and 3M_J planets
% duffell.PC.2015.10            % e grows to ~ 0.1 if planets can have initial e ~ 0.01
% ragusa.E.2018.03              % e grows to 0.14 in 3e5 orbits for 13M_J planets... 

In theory, the eccentricity evolution of a planet can be understood as the competition between Lindblad resonances, which excite the eccentricity, and corotation resonances, which typically damp the eccentricity (\citealt{goldreich.P.1980.10,goldreich.P.2003.03,ogilvie.GI.2003.04}). Massive gap-opening planets are thus more susceptible to eccentricity growth, as the damping corotation resonances, which are otherwise typically stronger than the Lindblad resonances, are weakened by the presence of a gap.
% goldreich.P.1980.10           % early (first?) work on Lindblad vs. corotation resonances
% goldreich.P.2003.03           % eccentricity growth through a finite-amplitude instability
% ogilvie.GI.2003.04            % theory works on the saturation of corotation resonances

While planet-disk interactions can excite the eccentricity of the planet, eccentric planets can also provide feedback on the structure of their disk. Unlike perturbing the disk with a single co-rotating pattern, as in the circular case, eccentric planet-disk interactions feature two intrinsic (azimuthal and temporal) periods. Therefore, the forcing can couple to a broader range of Lindblad and corotation resonances. The resulting gas response includes not only eccentric gaps, but also spiral density waves that exhibit detachments, bifurcations, and crossings (\citealt{goldreich.P.1980.10,hosseinbor.AP.2007.07,zhu.Z.2022.01,fairbairn.CW.2022.10}).
% goldreich.P.1980.10           % Lindblad and corotation resonances for eccentric planets
% hosseinbor.AP.2007.07         % eccentric planet create eccentric gaps 
% fairbairn.CW.2022.10          % eccentric potential decomposed into azimuthal and temporal harmonics
% zhu.Z.2022.01                 % complex spiral morphology for eccentric planets
This interaction can even excite a global eccentric disk mode associated with the outer 1:3 Lindblad resonance, once the planetary mass exceeds a threshold of a few Jupiter masses. Such disks may precess and produce observable gas signatures, including large-scale asymmetries in the density field, disturbed velocity fields, and persistent line-profile asymmetries (\citealt{kley.W.2006.02,regaly.Z.2010.11,teyssandier.J.2016.05,teyssandier.J.2017.06,baruteau.C.2021.05}).
% kley.W.2006.02                % eccentric disk mode excited by massive planets
% teyssandier.J.2016.05         % eccentric disk states can precess
% teyssandier.J.2017.06         % eccentric disk states can precess
% regaly.Z.2010.11              % eccentric disk has line-profile asymmetries
% baruteau.C.2021.05            % eccentric disk has large-scale asymmetries in gas density
Meanwhile, the planet's eccentricity also plays a critical role in regulating planetary accretion, affecting both the accretion rate and, in some cases, the rotational direction of the circumplanetary gas (\citealt{chen.YX.2022.11,li.YP.2023.12}).
% chen.YX.2022.11              % eccentric planets can make prograde CPD become retrograde
% li.YP.2023.12                % eccentric planets can boost the accretion rate

Although the response from the gas component of the disk to an eccentric planet has been studied extensively, the corresponding dust response remains comparatively underexplored. Moreover, existing studies either focus only on lower-mass planets that do not open deep gaps or only adopt two-dimensional (2D) models that do not capture any vertical structures, leaving 3D dust dynamics influenced by eccentric gap-opening planets largely unexamined (e.g., \citealt{ataiee.S.2013.05,chen.YX.2021.12}). 
% ataiee.S.2013.05            % 2D simulations of 10 MJ planets
% chen.YX.2021.12             % 2D simulations of saturn-mass planets and lower
In \cite{bi.J.2021.05}, hereafter \citetalias{bi.J.2021.05}, we investigated dust dynamics with circular gap-opening planets using 3D disk models. We found that the planet can puff up the dust layer at the gap edges by driving meridional gas circulation. The gas circulation can loft dynamically coupled dust grains to high disk elevations, producing a vertically extended dust distribution at the gap edges that may be observable. However, it remains unclear how this mechanism operates when the planet is eccentric. We therefore address this question in this work by carrying out 3D simulations with eccentric gap-opening planets embedded in protoplanetary disks. 

The paper is organized as follows. In Section~\ref{sec:methods}, we describe our simulation setup. In Section~\ref{sec:results}, we present the main results of our simulations. In Section~\ref{sec:discuss}, we discuss the implications of our findings. Finally, we summarize our conclusions in Section~\ref{sec:conclude}.

\section{Numerical Methods} \label{sec:methods}

Similar to \citetalias{bi.J.2021.05}, we considered a 3D global model of a protoplanetary disk consisting of both gas and dust, embedding a planet of mass $M_{\rm p}$, and orbiting a central star of mass $M_\star$. We did not include self-gravity, magnetic fields, planet migration, or planet accretion in our simulations.

In this paper, we use $\{r, \phi, \theta\}$ to denote the spherical radius, azimuthal angle, and polar angle, respectively. We also use $\{R, Z\}$ to denote the cylindrical radius and height. Both coordinates are centered on the star. The subscript ``0'' denotes azimuthally averaged quantities evaluated in the disk midplane at the reference radius $R_0$, while the superscript ``ini'' denotes the initial values of the disk model.

\subsection{Basic Equations and Model Setup} \label{sec:model}

The basic equations for our models are similar to those in \citetalias{bi.J.2021.05}. Therefore, we provide only a summary here and refer readers to that work for more detailed descriptions and justifications.

The volumetric density, pressure, and velocity vector of the gas are denoted by $\{\rho_{\rm g}, P, \textit{\textbf{V}}_{\rm g}\}$. We assumed a vertically isothermal equation of state for the gas $P = \rho_{\rm g}c_{\rm s}^2$, where $c_{\rm s}$ is the sound speed. Following \citet{chiang.EI.1997.11}, we adopted a time-independent, axisymmetric, flared disk geometry with an aspect ratio of $h_{\rm g} = H_{\rm g}/R = h_{\rm g0}(R/R_0)^{2/7}$. In the above, $H_{\rm g} = c_{\rm s}/\Omega_{\rm K}$ is the gas scale height, $\Omega_{\rm K} = \sqrt{GM_\star/R^3}$ is the Keplerian angular velocity, and $G$ is the gravitational constant. 
We adopt $h_{\rm g0} = 0.05$ and a constant kinematic viscosity $\nu = 10^{-5}R_0^2\Omega_{\rm K0}$, corresponding to an alpha viscosity parameter of $\alpha_0 = \nu/(c_{\rm s0}H_{\rm g0}) = 4\times10^{-3}$ \citep{shakura.NI.1973.01}. This relatively high viscosity is chosen to suppress the vertical shear instability, which can otherwise puff up the dust layer and thereby contaminate the dust substructures driven by the planet-disk interactions of interest \citep{nelson.RP.2013.11,stoll.MHR.2014.12,lin.MK.2019.06,flock.M.2020.07}.

The volumetric density and velocity vector of the dust are denoted by $\{\rho_{\rm d}, \textit{\textbf{V}}_{\rm d}\}$. The dust is modeled as a pressureless fluid. Dust feedback onto the gas is included, whereas dust diffusion and dust collisions are neglected. The aerodynamic coupling between gas and dust is parameterized by the Stokes number ${\rm St} = \tau_{\rm s}\Omega_{\rm K}$. We considered one dust species with ${\rm St}_0 = 10^{-3}$, whose grain size is fixed throughout the disk. Here, ${\rm St}_0$ denotes the reference Stokes number in an unperturbed disk without the planet-opened gap at $R_0$, and is therefore equivalent to ${\rm St}_0^{\rm ini}$. We assume that dust grains are in the Epstein regime \citep{epstein.PS.1924.06}, in which case the particle stopping time $\tau_{\rm s}$ is given by
\begin{equation}
\tau_{\rm s} = \frac{\rho_{\rm g0}^{\rm ini}}{\rho_{\rm g}}\frac{c_{\rm s0}}{c_{\rm s}}\frac{{\rm St}_0}{\Omega_{\rm K0}}.
\end{equation}

The hydrodynamic equations for gas and dust are given by 
\begin{align} 
    &\frac{\partial \rho_{\rm g}}{\partial t} + \nabla \cdot (\rho_{\rm g} \textit{\textbf{V}}_{\rm g}) = 0, \\
    &\begin{aligned}
        \frac{\partial \textit{\textbf{V}}_{\rm g}}{\partial t} + \textit{\textbf{V}}_{\rm g} \cdot \nabla \textit{\textbf{V}}_{\rm g} = &- \frac{1}{\rho_{\rm g}} \nabla P - \nabla \Phi \\
        &+ \frac{\epsilon}{\tau_{\rm s}}(\textit{\textbf{V}}_{\rm d} - \textit{\textbf{V}}_{\rm g}) + \frac{1}{\rho_{\rm g}} \nabla \cdot \mathcal{T},
    \end{aligned} \\
    &\frac{\partial \rho_{\rm d}}{\partial t} + \nabla \cdot (\rho_{\rm d} \textit{\textbf{V}}_{\rm d}) = 0, \\ 
    &\frac{\partial \textit{\textbf{V}}_{\rm d}}{\partial t} + \textit{\textbf{V}}_{\rm d} \cdot \nabla \textit{\textbf{V}}_{\rm d} =  - \nabla \Phi - \frac{1}{\tau_{\rm s}}(\textit{\textbf{V}}_{\rm d} - \textit{\textbf{V}}_{\rm g}).
\end{align}
Here, $\Phi$ is the net gravitational potential from the star, the planet, including the indirect star--planet interaction term \citep{regaly.Z.2017.05,crida.A.2025.07}. $\epsilon\equiv\rho_{\rm d}/\rho_{\rm g}$ is the local dust-to-gas density ratio, and $\mathcal{T}(\nu)$ is the viscous stress tensor.

The planet is placed on a fixed orbit in the disk midplane, with semi-major axis $a_{\rm p} = R_0$ and eccentricity $e_{\rm p}$. Four values of the eccentricity are explored, with $e_{\rm p} \in \{0,\,0.05,\,0.1,\,0.2\}$. We consider two planet masses with $M_{\rm p} \in \{3\times10^{-4}\,M_\star,\,1\times10^{-3}\,M_\star\}$. These correspond to a Saturn-like planet with $q / h_{\rm g0}^3 = 2.4$ and a Jupiter-like planet with $q / h_{\rm g0}^3 = 8.0$, respectively, where $q = M_{\rm p} / M_\star$ is the planet-to-star mass ratio. To avoid singularities, the gravitational potential of the planet is smoothed over a length scale of $0.1\,H_{\rm g0}$ \citep{muller.TWA.2012.05}.

\subsection{Initialization} \label{sec:initial}

The gas density is initialized with an azimuthally symmetric profile given by
\begin{equation}
    \rho_{\rm g}^{\rm ini} = \rho_{\rm g0}^{\rm ini}\left(\frac{R}{R_0}\right)^{-1/2}\times \exp\left[-\frac{GM_\star}{c_{\rm s}^2}\left(\frac{1}{r}-\frac{1}{R}\right)\right],
\end{equation}
where the normalization factor $\rho_{\rm g0}^{\rm ini}$ can be arbitrary for our non-self-gravitating disk. Dust is initialized through the dust-to-gas ratio $\epsilon^{\rm ini} \equiv \rho_{\rm d}^{\rm ini}/\rho_{\rm g}^{\rm ini} = \epsilon_0^{\rm ini}\times\delta_R\times\delta_Z$. The factor $\delta_R$ is a radial taper introduced so that $\rho_{\rm d}^{\rm ini}$ gradually approaches zero near the radial boundaries, while $\delta_Z$ is a vertical taper ensuring that $\rho_{\rm d}^{\rm ini}$ follows a pre-settled, vertically Gaussian profile with a dust scale height $H_{\rm d}$. We adopted $\epsilon_0^{\rm ini} = 0.01$ and $H_{\rm d}^{\rm ini} = 0.2\,H_{\rm g}^{\rm ini}$, corresponding to a total dust-to-gas mass ratio of $\sim$0.2\%. This relatively low dust-to-gas mass ratio is chosen to avoid attributing the weaker dust puff-up features in less eccentric cases, as shown in Section~\ref{sec:dust_disk}, to stronger dust feedback onto the gas, which could otherwise arise because the dust rings are better maintained in those cases.

The azimuthal velocity of gas and dust are initialized to $V_{\rm g,\phi}^{\rm ini} = \sqrt{1 - 2\eta}R\Omega_{\rm K}(R)$ and $V_{\rm d,\phi}^{\rm ini} = r\Omega_{\rm K}(r)$, where $\eta$ is a dimensionless measurement of the radial gradient of gas pressure. The radial and polar velocities of gas and dust are all initialized to zero. 

The planet is always initialized at apoapsis. To mitigate transient effects, the planet mass is gradually increased from zero to $M_{\rm p}$ over the first 10 orbits.

\subsection{Numerical Setup} \label{sec:numerical}

The model is evolved by \texttt{FARGO3D} \citep{benitez-llambay.P.2016.03,benitez-llambay.P.2019.04}. The computational domain extends from $0.4R_0$ to $4.0R_0$ in $r$, covers the full $2\pi$ in $\phi$, and extends from the disk midplane to $\pm3h_{\rm g0}$ in $\theta$. The grids are logarithmically spaced in $r$ and uniformly spaced in $\theta$ and $\phi$. The grid resolutions are $N_\theta\times N_r\times N_\phi = 90\times450\times630$, corresponding to $\sim$15, 10, and 5 cells per $H_{\rm g0}$. For the Saturn-like planet ($R_{\rm H} \lesssim H_{\rm g0}$), these resolutions correspond to $\sim$14, 9, and 5 cells per $R_{\rm H}$, while for the Jupiter-like planet ($R_{\rm H} \gtrsim H_{\rm g0}$), they correspond to $\sim$21, 14, and 7 cells per $R_{\rm H}$. Here,
\begin{equation}
    R_{\rm H} = a_{\rm p}\sqrt[3]{\frac{q}{3(1 + q)}}
\end{equation}
is the Hill radius of the planet.

Periodic conditions are applied at $\phi$ boundaries. $\rho_{\rm d}$ is symmetric at $r$ and $\theta$ boundaries, while $\rho_{\rm g}$ and $V_\phi$ are extrapolated there. $V_\theta$ is antisymmetric at $\theta$ boundaries and symmetric at $r$ boundaries, while $V_r$ is antisymmetric at $r$ boundaries and symmetric at $\theta$ boundaries, except that the inner radial boundary is open for mass loss of dust. Wave-killing conditions are applied to the gas at $r$ boundaries within a margin of $\sim$10\% of the boundary radius. The simulations are evolved for $1000\,T_0$, where $T_0 = 2\pi\Omega_{\rm K0}^{-1}$ is the reference orbital period.

\section{Results} \label{sec:results}

Although the primary focus of this study is the vertical puff-up of the dust layer in the radial--vertical plane, the use of azimuthally averaged analyses, similar to those in \citetalias{bi.J.2021.05}, requires further justification when the disk becomes increasingly eccentric as a result of the planet's eccentricity (e.g., \citealt{kley.W.2006.02}). We therefore first examine the eccentricity of the gas component in Section~\ref{sec:gas_disk}. We then assess the reliability of azimuthally averaged profiles for characterizing the disk structure in the eccentric cases in Section~\ref{sec:azi_avg}. Following these preliminary but necessary analyses, we present the morphology of the dust component in the radial--vertical plane in Section~\ref{sec:dust_disk}.

\subsection{Morphology of the Gas Component} \label{sec:gas_disk}

\begin{figure*}
\centering
\includegraphics[width = \textwidth]{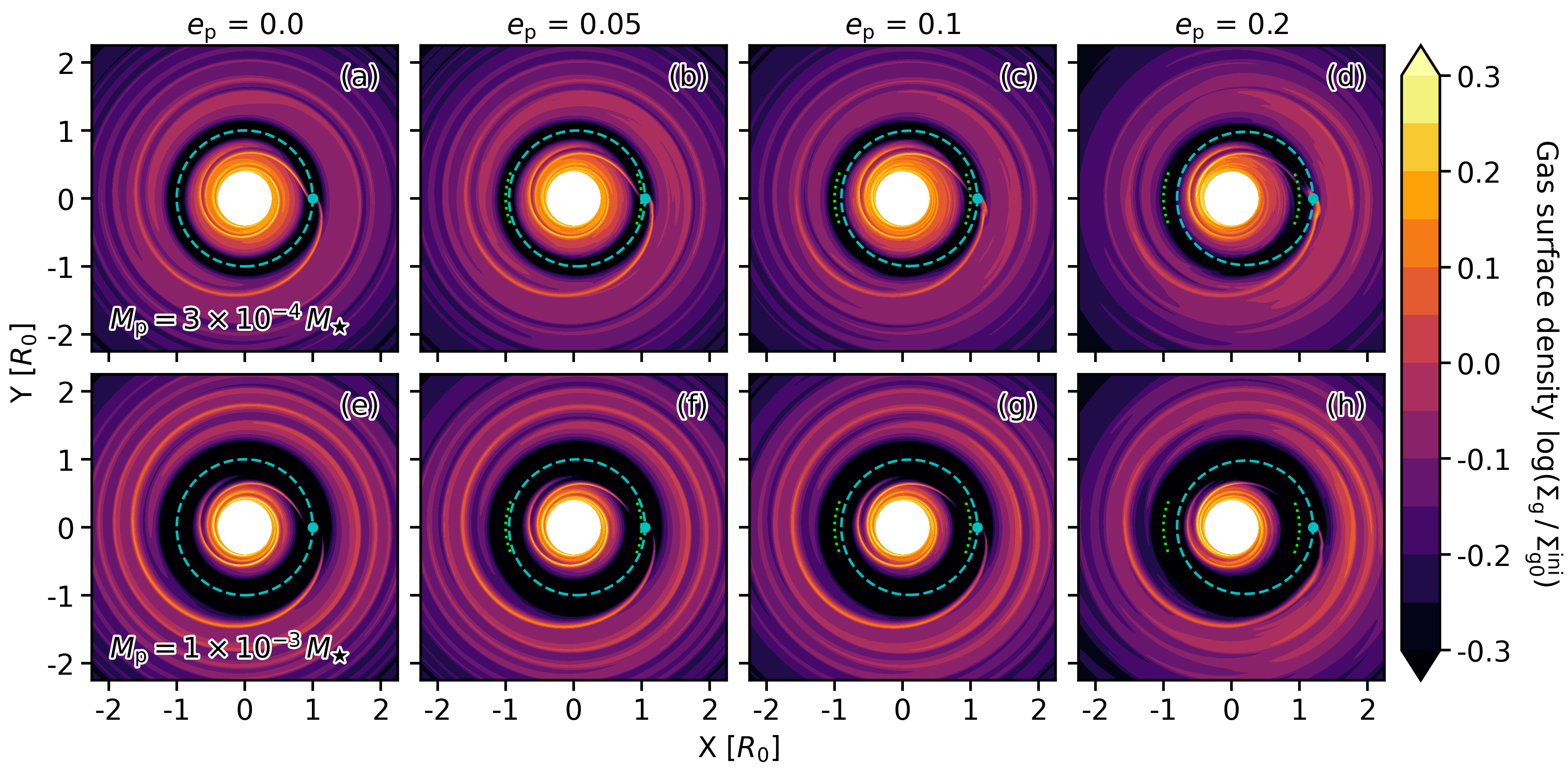}
\figcaption{Gas surface density at $t = 1000\,T_0$ in models with different planet masses $M_{\rm p}$ and orbital eccentricities $e_{\rm p}$. Panels in the same row share the same $M_{\rm p}$, while panels in the same column share the same $e_{\rm p}$. The planet is located on the $+x$-axis at apoapsis and is marked by a cyan dot. The dashed cyan trajectory marks the planet's orbit, and the dotted green trajectories near the planet's periapsis and apoapsis mark the corresponding circular orbits with the same semi-major axis. 
\label{fig:xy_sigmag}}
\end{figure*}

\begin{figure*}
\centering
\includegraphics[width = \textwidth]{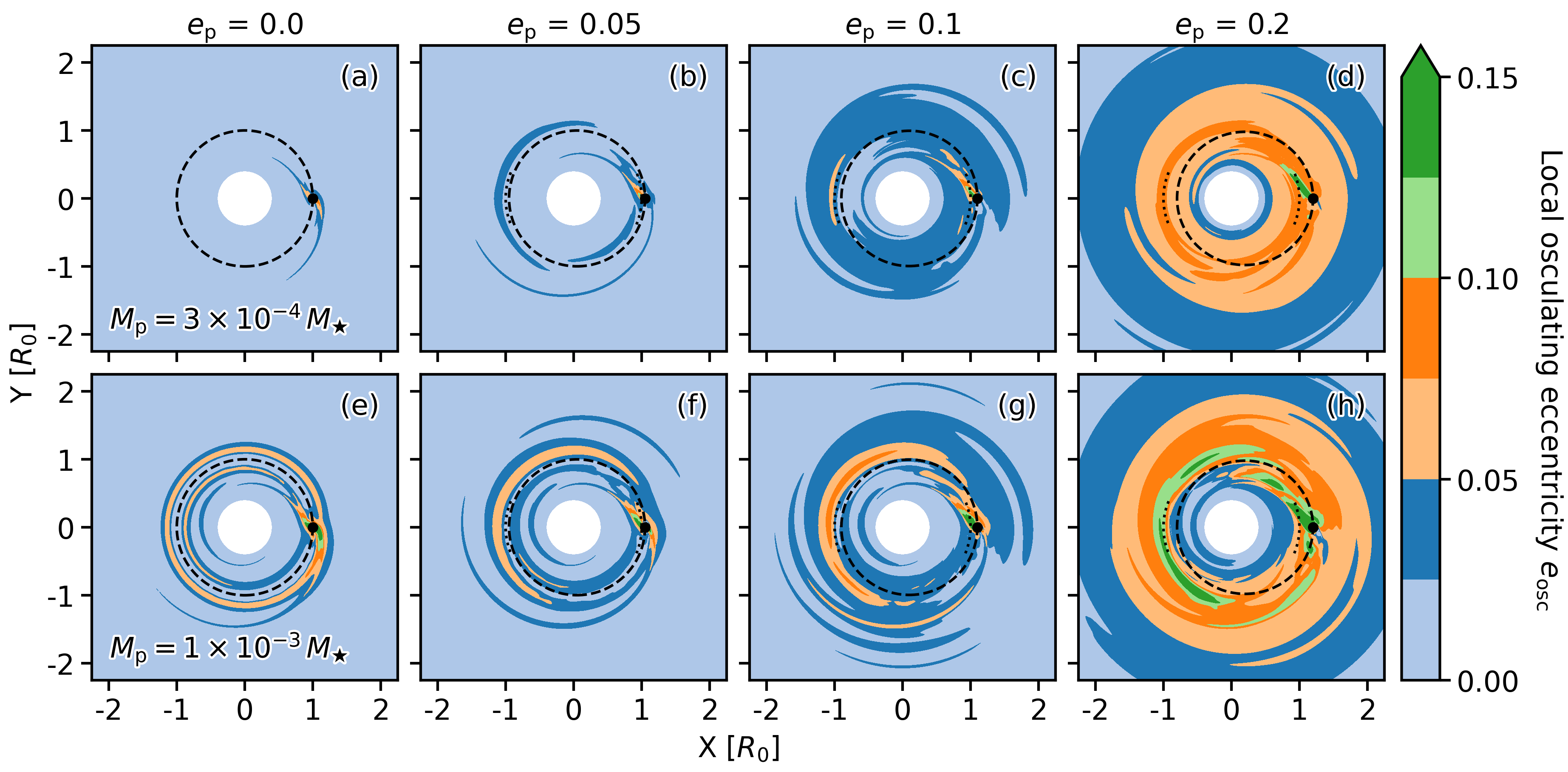}
\figcaption{Local osculating eccentricity of the gas disk $e_{\rm osc}$ at $t = 1000\,T_0$ in models with different planet masses $M_{\rm p}$ and orbital eccentricities $e_{\rm p}$. Panels in the same row share the same $M_{\rm p}$, while panels in the same column share the same $e_{\rm p}$. The planet is located on the $+x$-axis at apoapsis and is marked by a black dot. The dashed trajectory marks the planet's orbit, and the dotted trajectories near the planet's periapsis and apoapsis mark the corresponding circular orbits with the same semi-major axis.
\label{fig:xy_eccosc}}
\end{figure*}

Figure~\ref{fig:xy_sigmag} shows the gas surface density in models with different planet masses and orbital eccentricities. Qualitatively consistent with previous 2D simulations by \citet{dangelo.G.2006.12}, we find that the inner spirals retain a regular morphology in most cases, whereas the outer spirals become increasingly distorted as the planet eccentricity increases. In agreement with \citet{zhu.Z.2022.01}, we also identify spirals that are detached from the eccentric planet, as well as spirals that bifurcate into two branches, most notably in panel~(d) of Figure~\ref{fig:xy_sigmag}.

Figure~\ref{fig:xy_eccosc} shows the local \textit{osculating} eccentricity of the gas, $e_{\rm osc} = |\bm{e}|$. Following \citet{dangelo.G.2006.12}, we computed the eccentricity vector $\bm{e}$ by treating each gas cell as a pressureless test particle with the same instantaneous position and mass-weighted, vertically averaged velocity, measured with respect to the barycenter of the planet-star system:
\begin{equation} \label{eq:osculating}
\bm{e} = \frac{\bm{v^\prime}\times(\bm{r^\prime}\times\bm{v^\prime})}{G(M_\star + M_{\rm p})} - \frac{\bm{r^\prime}}{|\bm{r^\prime}|},
\end{equation}
where $\bm{r^\prime}$ and $\bm{v^\prime}$ are the position and velocity vectors of the gas cell relative to the barycenter.

The planet excites the strongest disk eccentricity in its immediate vicinity, where the gas motion is associated with planetary wakes. A more eccentric planet excites higher disk eccentricity, and increasing the planet mass further amplifies this excitation. Nevertheless, the disk eccentricity typically remains $e_{\rm osc}\lesssim 0.5\,e_{\rm p}$ in our parameter space. Although the Jupiter-like planet can drive a moderate eccentricity at the gap edges, with $e_{\rm osc}\sim0.05$, in the circular-orbit case (panel~(e) of Figure~\ref{fig:xy_eccosc}), we caution that this diagnostic may not accurately reflect the gas dynamics there. This is because Equation~\ref{eq:osculating} neglects gas pressure, shocks, and viscous stress in the orbital element construction. Therefore, $e_{\rm osc}$ may include contributions from non-eccentric motion and should be interpreted only as a local kinematic diagnostic, rather than the true disk eccentricity.

\begin{figure}[htbp]
\centering
\includegraphics[width = \columnwidth]{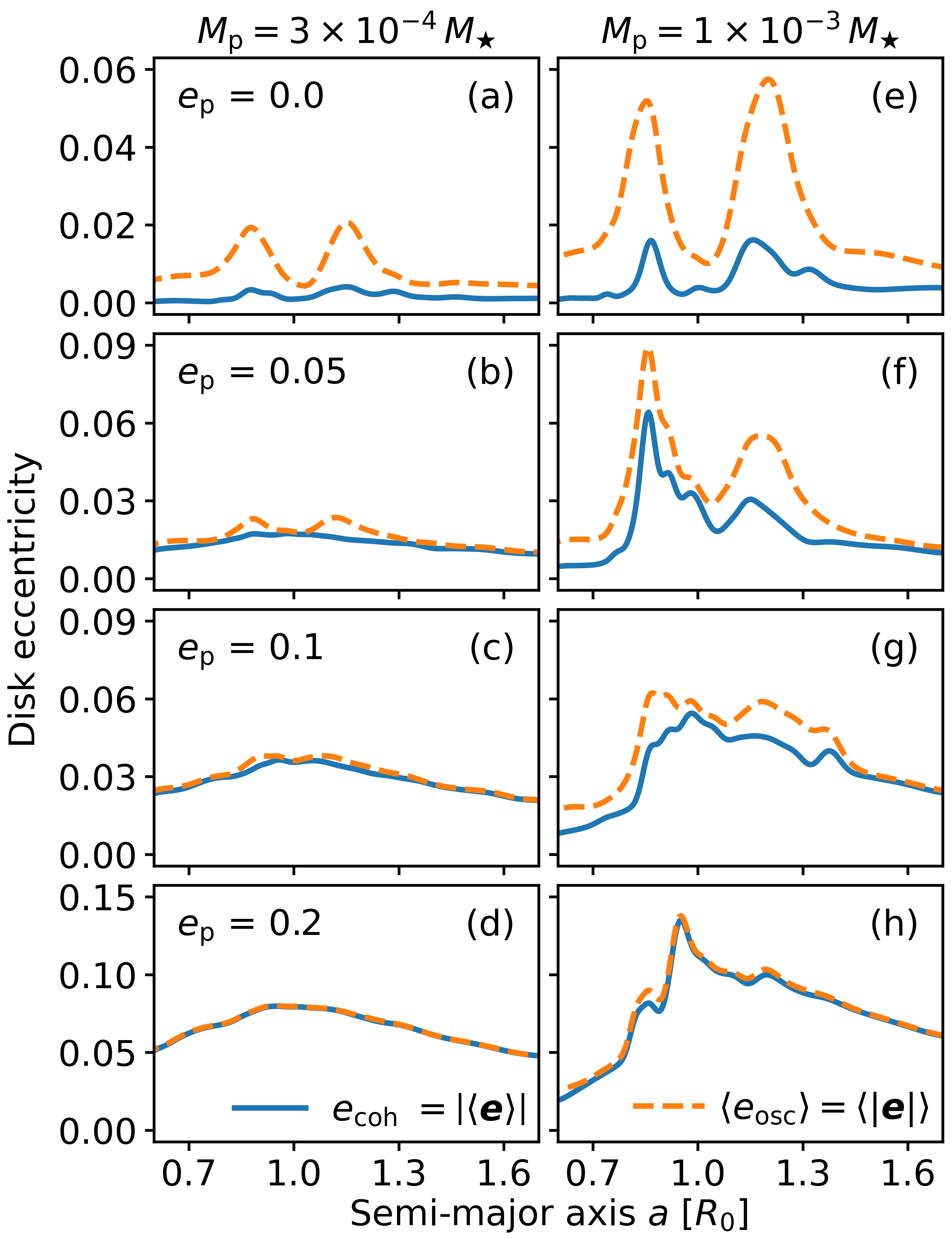}
\figcaption{Coherent eccentricity $e_{\rm coh}$ and local osculating eccentricity averaged within semi-major axis bins $\langle e_{\rm osc} \rangle$ of the gas disk at $t = 1000\,T_0$ in models with different planet masses $M_{\rm p}$ and orbital eccentricities $e_{\rm p}$. The legend for the line styles is provided in the bottom panels. Panels in the same row share the same $e_{\rm p}$, while panels in the same column share the same $M_{\rm p}$. The region within a distance $H_{\rm g0}$ from the planet is masked out in the calculations, and a one-dimensional Gaussian convolution with a radius-dependent full width at half maximum of $H_{\rm g}(R)$ is applied to the profiles to remove spiky features introduced by the velocity field.
\label{fig:ry_ecccoh}}
\end{figure}

Following the procedure described in \citet{teyssandier.J.2017.06}, we calculated the eccentricity $e_{\rm coh}$, which more accurately illustrates the \textit{coherent} eccentric motion in the disk. Note that Equation~\ref{eq:osculating} can be written in complex form as
\begin{equation}
    \bm{e} = e_{\rm osc}\times\exp(i\varpi),
\end{equation}
where $\varpi$ is the longitude of periapsis. Then $e_{\rm coh} = |\langle\bm{e}\rangle|$, where $\langle...\rangle$ denotes the mass-weighted averaging within each semi-major axis bin, and the semi-major axis $a$ is obtained from the vis-viva equation
\begin{equation}
    |\bm{v^\prime}|^2 = G(M_\star + M_{\rm p})\left(\frac{2}{|\bm{r^\prime}|} - \frac{1}{a}\right).
\end{equation} 
Unlike $\langle e_{\rm osc} \rangle = \langle|\bm{e}|\rangle$, $e_{\rm coh}$ allows eccentric motions with different $\varpi$ to cancel out, thereby isolating coherent eccentric modes with an aligned orientation. 

The comparisons between $e_{\rm coh}$ and $\langle e_{\rm osc} \rangle$ are shown in Figure~\ref{fig:ry_ecccoh}. Since $e_{\rm coh} \ll \langle e_{\rm osc} \rangle$ in panels~(a) and (e), the eccentricity signals at the gap edges in the circular-orbit cases, especially for the Jupiter-like planet in panel~(e) of Figure~\ref{fig:xy_eccosc}, are evidently dominated by pressure-gradient effects rather than coherent eccentric motion. In contrast, $e_{\rm coh}$ and $\langle e_{\rm osc} \rangle$ become increasingly consistent as the planet eccentricity increases (panels~(b)--(d) and (f)--(h) of Figure~\ref{fig:ry_ecccoh}), indicating that Figure~\ref{fig:xy_eccosc}, especially panels~(d) and (h), indeed captures eccentricity excitation by the eccentric planet. 

\subsection{How Wrong Are Azimuthally Averaged Profiles?} \label{sec:azi_avg}

Azimuthally averaged diagnostics are useful tools for characterizing radial--vertical structures in the disk, especially for our 3D models, where the full 3D structure is nontrivial to visualize. However, this approach relies on the assumption that the disk remains approximately axisymmetric. As shown in Section~\ref{sec:gas_disk}, the disk morphology becomes increasingly non-axisymmetric as the planet eccentricity increases. It is therefore important to examine how well azimuthally averaged profiles represent the disk structure in such cases.

\begin{figure*}
\centering
\includegraphics[width = \textwidth]{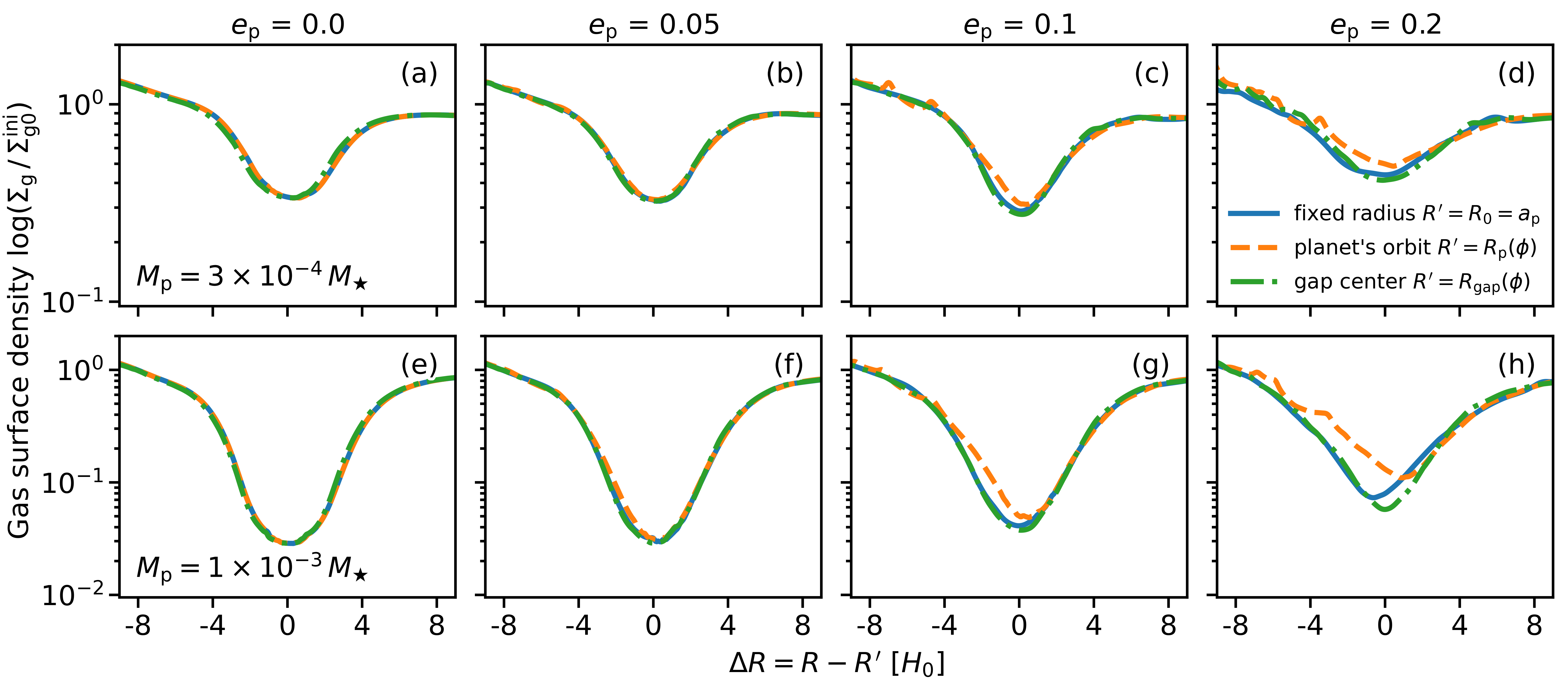}
\figcaption{Gas surface density at $t = 1000\,T_0$ in models with different planet masses $M_{\rm p}$ and orbital eccentricities $e_{\rm p}$. Different curves in each panel represent azimuthal averaging with respect to different radial locations. The legend for the line styles is provided in the bottom panels. Panels in the same row share the same $M_{\rm p}$, while panels in the same column share the same $e_{\rm p}$. The region within a distance $H_{\rm g0}$ from the planet is masked out in the calculations.
\label{fig:ry_sigmag}}
\end{figure*}

One important diagnostic is the planet-opened gap in the profile of $\Sigma_{\rm g}$. Figure~\ref{fig:ry_sigmag} compares profiles obtained by azimuthal averaging with respect to three reference radii: the semi-major axis of the planet $R_0$, the planet's orbital radius $R_{\rm p}(\phi)$, and the center radius of the planet-opened gap $R_{\rm gap}(\phi)$. In principle, using $R_{\rm gap}(\phi)$ should provide the most accurate representation of the gap, but this reference is also the least straightforward to determine. Averaging with respect to $R_0$ is the standard and simplest choice, although it can become inaccurate when the gap is eccentric. Averaging with respect to $R_{\rm p}(\phi)$ provides an intermediate option between these two limits, but it can also be inaccurate when the planet's orbit deviates substantially from the gap center, as can occur in moderately eccentric cases. To obtain the profile with respect to $R_{\rm p}(\phi)$, we radially shifted each azimuthal slice by $\delta_R = R_{\rm p}(\phi)-R_0$ using cubic interpolation. For the profile with respect to $R_{\rm gap}(\phi)$, we first identified, in each azimuthal slice, the radial region corresponding to the lowest 10\% of $\Sigma_{\rm g}$, and then used the center of this region as $R_{\rm gap}(\phi)$ to shift the azimuthal slices.

Figure~\ref{fig:ry_sigmag} shows that, using the profile constructed with respect to $R_{\rm gap}(\phi)$ as a benchmark, standard azimuthal averaging still captures the gap location and depth reasonably well, even in the most eccentric cases (panels~(d) and (h)). In contrast, the profile constructed with respect to $R_{\rm p}(\phi)$ fails to accurately recover the gap structure in the eccentric cases. We therefore conclude that azimuthal averaging with respect to $R_0$ remains a reliable approximation for the models in our chosen parameter space, and we adopt this standard procedure in the subsequent analyses.

\subsection{Morphology of the Dust Component} \label{sec:dust_disk}

\begin{figure*}
\centering
\includegraphics[width = \textwidth]{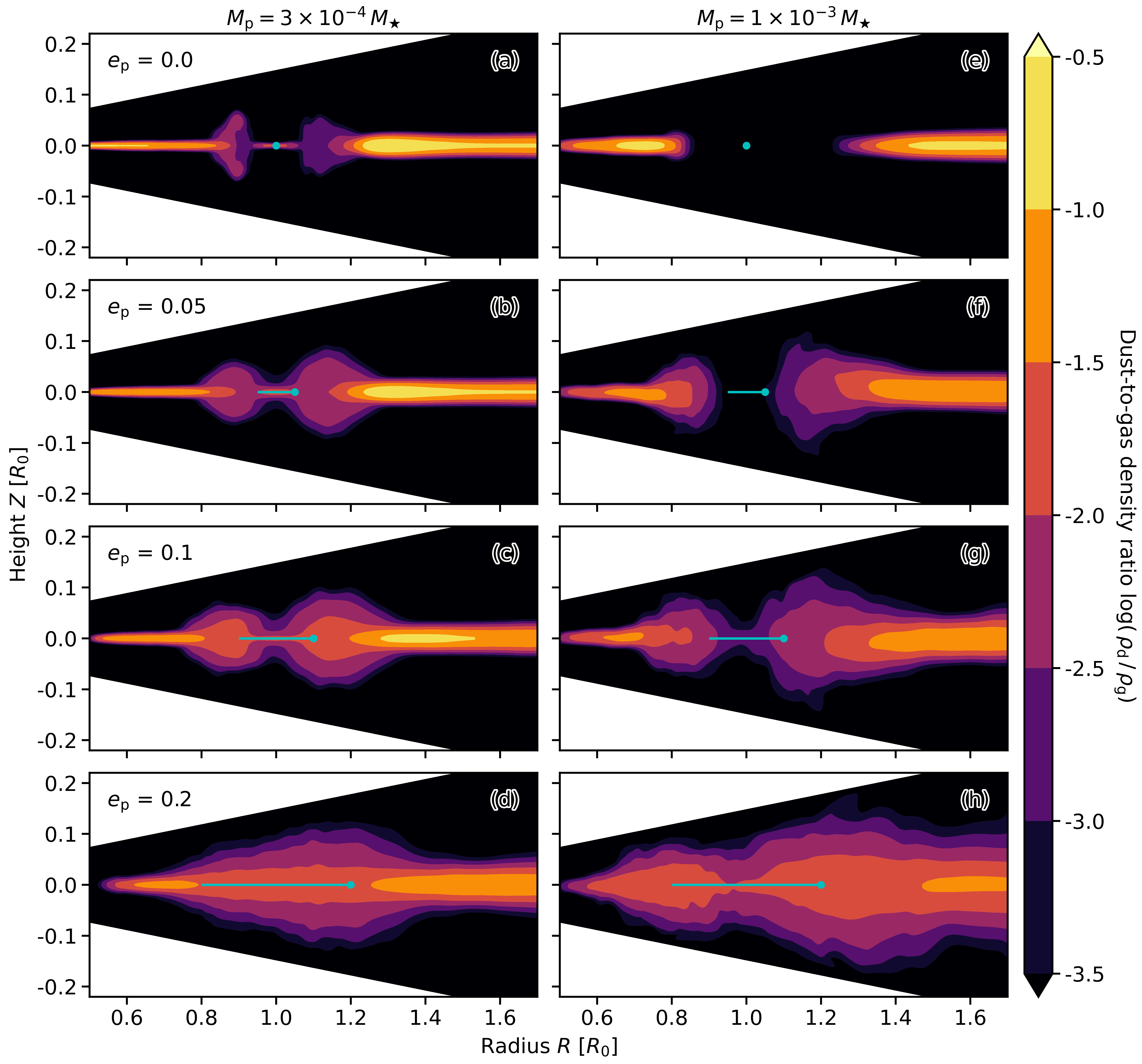}
\figcaption{Azimuthally averaged dust-to-gas volumetric density ratio at $t = 1000\,T_0$ in models with different planet masses $M_{\rm p}$ and orbital eccentricities $e_{\rm p}$. Panels in the same row share the same $e_{\rm p}$, while panels in the same column share the same $M_{\rm p}$. The planet is located at apoapsis and is marked by a cyan dot. The cyan line marks the radial range traversed by the planet over its orbit.
\label{fig:rz_rhod2g}}
\end{figure*}

\begin{figure*}
\centering
\includegraphics[width = \textwidth]{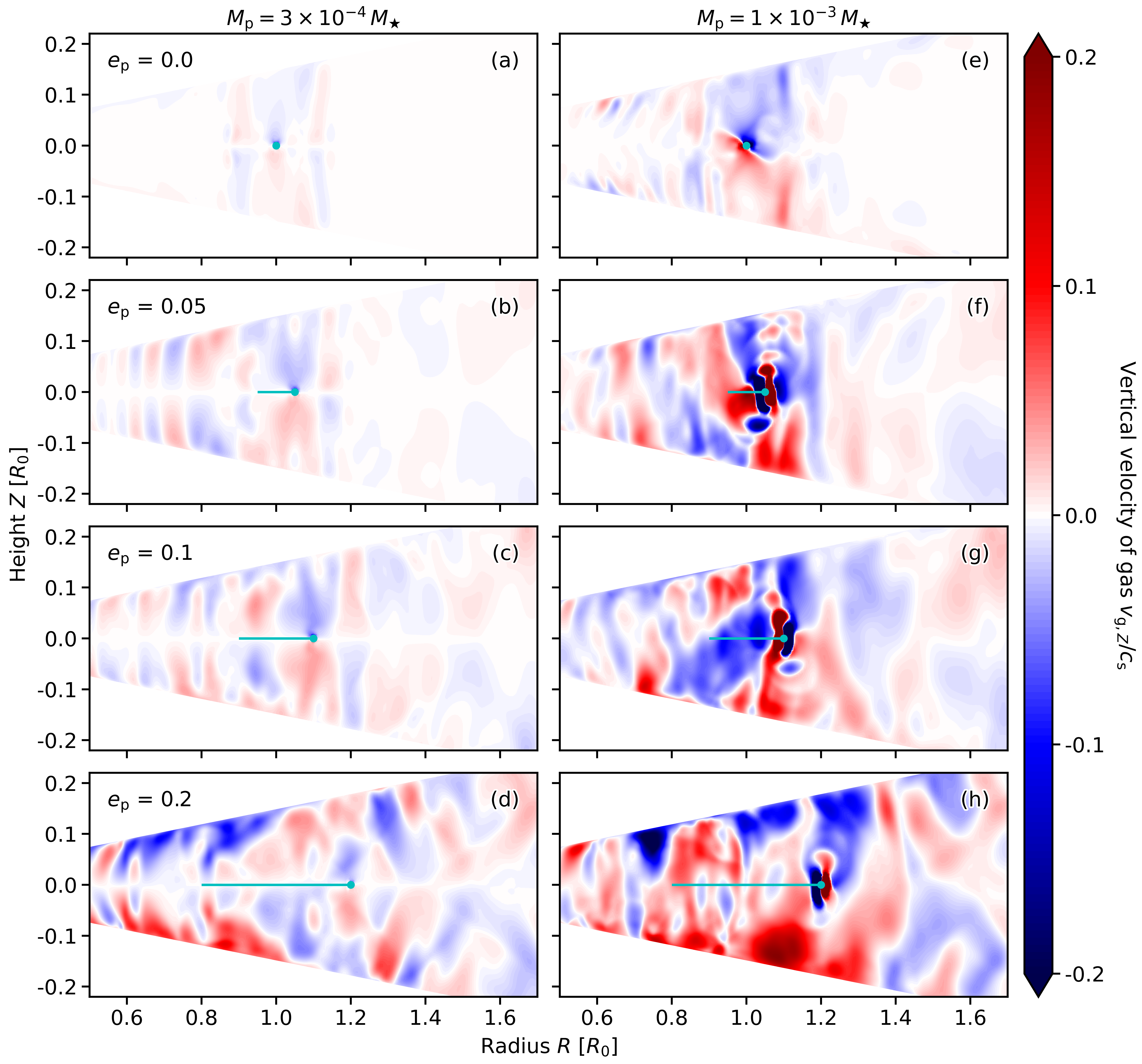}
\figcaption{Mass-weighted, azimuthally averaged vertical velocity of the gas at $t = 1000\,T_0$ in models with different planet masses $M_{\rm p}$ and orbital eccentricities $e_{\rm p}$. Panels in the same row share the same $e_{\rm p}$, while panels in the same column share the same $M_{\rm p}$. The planet is located at apoapsis and is marked by a cyan dot. The cyan line marks the radial range traversed by the planet over its orbit.
\label{fig:rz_gas_vz}}
\end{figure*}

Figure~\ref{fig:rz_rhod2g} shows the azimuthally averaged dust-to-gas density ratio in models with different planet masses and eccentricities. To assess the robustness of azimuthal averaging for this diagnostic, we also present this quantity plotted for individual azimuthal slices in Appendix~\ref{sec:d2gazi}. 

In \citetalias{bi.J.2021.05}, we used a model similar to that shown in panel~(a) of Figure~\ref{fig:rz_rhod2g} to demonstrate that planet-induced meridional gas circulation can puff up the dust layer at the gap edges. The comparison between the upper and lower panels of Figure~\ref{fig:rz_rhod2g} echoes our previous finding that more massive planets generally drive a stronger puff-up of the dust layer. However, this trend does not necessarily continue indefinitely with increasing planet mass. For planets more massive than the thermal mass, whose Hill radii exceed the disk scale height, the gap in the circular-orbit case can become so strongly depleted that dust grains with ${\rm St}\sim 10^{-3}$ are completely cleared from the gap region by the radial component of the meridional circulation \citep{bi.J.2024.08}. In such cases, little dust remains near the gap edges to be vertically lofted, and the puff-up signature can therefore become weak or even absent, as in panel~(e) of Figure~\ref{fig:rz_rhod2g}. 

The comparison from top to bottom in Figure~\ref{fig:rz_rhod2g} shows that planet eccentricity can greatly amplify the puff-up of the dust layer at the gap edges. This amplification is driven by the stronger meridional gas circulation excited by eccentric planets, as shown in Figure~\ref{fig:rz_gas_vz}, which loft dust grains to higher disk elevations. The trend of amplified gas flows at a higher planet eccentricity is evident in both planet mass regimes. In the Saturn-like cases, where the planet is below the thermal mass, gap opening remains relatively gentle and the vertical gas motion is approximately symmetric about the disk midplane (left panels of Figure~\ref{fig:rz_gas_vz}). In the Jupiter-like cases, where the planet exceeds the thermal mass, gap opening becomes more nonlinear and the vertical gas motion develops a clear midplane asymmetry (right panels of Figure~\ref{fig:rz_gas_vz}).

These stronger meridional flows at higher planet eccentricities may arise because gas streamlines in an eccentric disk undergo periodic compression and expansion, which can continuously perturb vertical hydrostatic equilibrium and thereby facilitate additional vertical gas motions. In inviscid environments, such periodic forcing may further trigger a parametric instability that eventually saturates into hydrodynamic turbulence \citep{pierens.A.2020.08,pierens.A.2021.12}. However, we do not have sufficient evidence to directly attribute the enhanced meridional flows in our models to this parametric instability, especially given the relatively high viscosity adopted in our simulations. We therefore conclude only that the stronger meridional flows in the eccentric cases are likely associated with the eccentric disk mode excited by the eccentric planet, and we leave a more detailed investigation of the underlying mechanism to future work.

\begin{figure*}
\centering
\includegraphics[width = \textwidth]{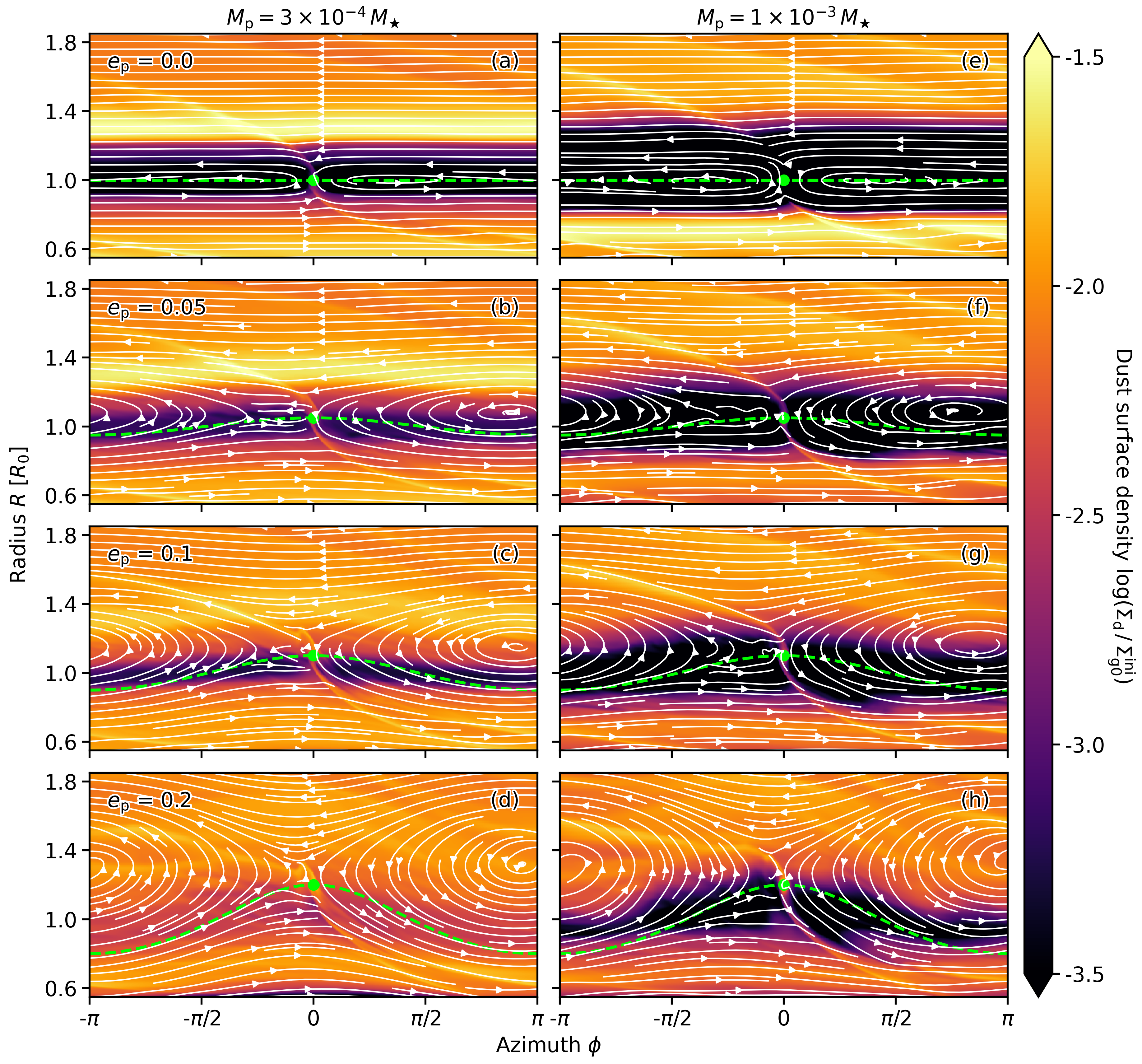}
\figcaption{Dust surface density at $t = 1000\,T_0$ in models with different planet masses $M_{\rm p}$ and orbital eccentricities $e_{\rm p}$. The white streamlines trace the mass-weighted non-vertical dust velocity field in the frame co-rotating with the planet. Panels in the same row share the same $e_{\rm p}$, while panels in the same column share the same $M_{\rm p}$. The planet is located at apoapsis and is marked by a green dot. The dashed green trajectory marks the planet's orbit.
\label{fig:ra_sigmad}}
\end{figure*}

In addition to the enhanced dust puff-up, we find that increasing planet eccentricity makes the gap in the dust-to-gas ratio much less pronounced, as shown in Figure~\ref{fig:rz_rhod2g}. This effect is unlikely to be an artifact of azimuthal averaging, in which different parts of an eccentric gap are stacked without radial alignment, because the individual azimuthal slices shown in Appendix~\ref{sec:d2gazi} also exhibit weaker gap contrasts. Moreover, this effect is unlikely to be restricted to the dust-to-gas ratio. Since the gas in the gap is already less depleted in the eccentric cases, as shown in Figure~\ref{fig:ry_sigmag}, a less pronounced gap in the dust-to-gas ratio implies an even less pronounced gap in the density of dust.

To understand this phenomenon, we plot the dust surface density together with streamlines of the mass-weighted velocity of dust in Figure~\ref{fig:ra_sigmad}. We find that increasing planet eccentricity not only makes the gas gap shallower, as shown in Figure~\ref{fig:ry_sigmag}, but also progressively fills in the dust gap. This occurs because the planet's horseshoe region becomes much wider in the radial direction as the planet eccentricity increases. As a result, the dust streamlines are no longer parallel to the azimuthal direction, as in panels~(a) and (e) of Figure~\ref{fig:xy_sigmad}, but instead facilitate active radial transport between the gap and the surrounding dust disk, thereby reducing the gap contrast. Consequently, the dust rings at the gap edges become less well defined and can even disappear in the most eccentric Saturn-like case (panel~(d) of Figure~\ref{fig:ra_sigmad}).

\section{Discussion} \label{sec:discuss}

\subsection{How Eccentric Are the Dust Rings?}

\begin{figure*}
\centering
\includegraphics[width = \textwidth]{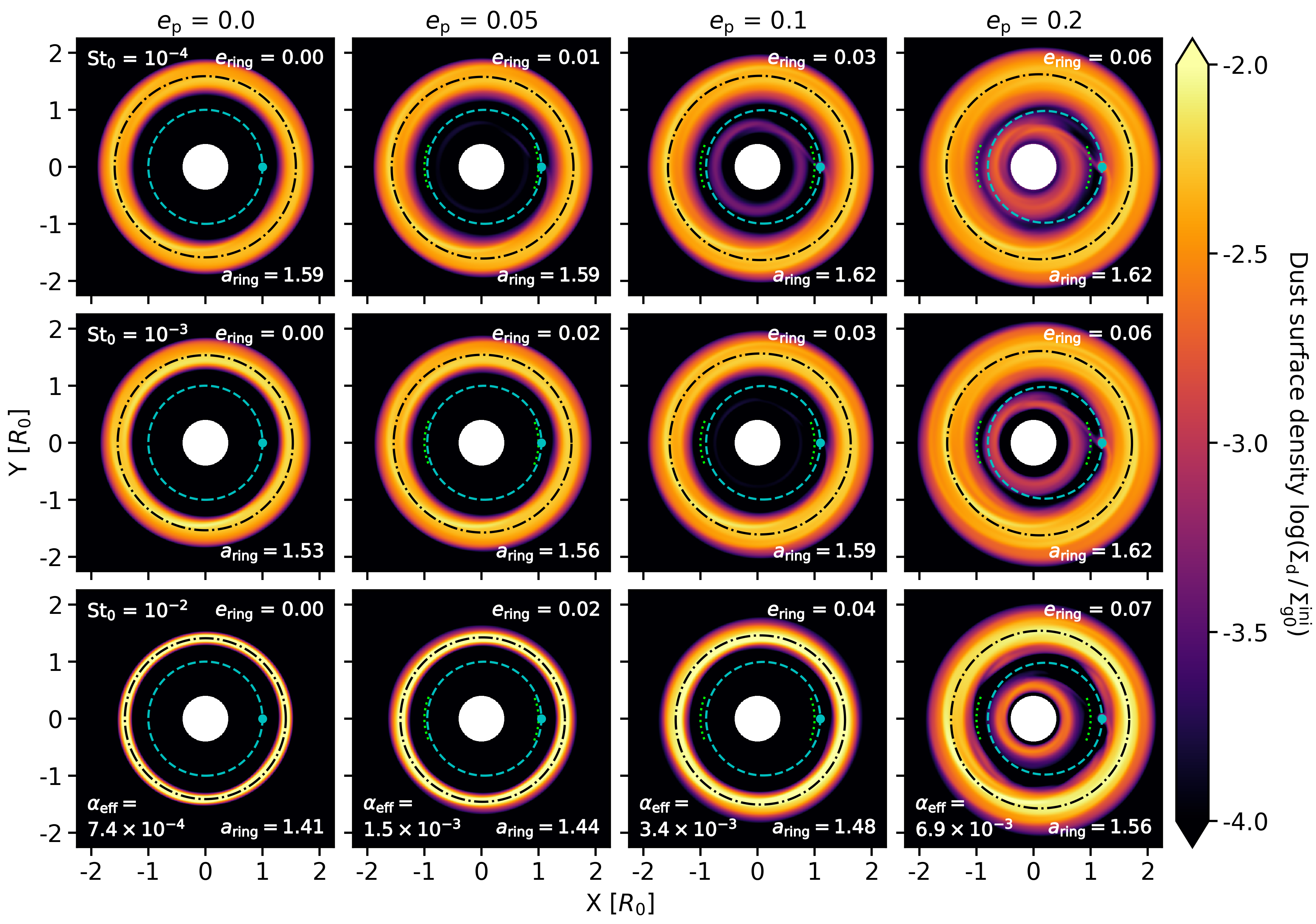}
\figcaption{Dust surface density of grains with different Stokes numbers ${\rm St}_0$ at $t = 1000\,T_0$ from the ``dust ring'' models with a Jupiter-like planet $M_{\rm p} = 10^{-3}M_\star$ and different orbital eccentricities $e_{\rm p}$. Panels in the same row share the same ${\rm St}_0$, while panels in the same column share the same $e_{\rm p}$. The planet is located on the $+x$-axis at apoapsis and is marked by a cyan dot. The dashed cyan trajectory marks the planet's orbit, and the dotted green trajectories near the planet's periapsis and apoapsis mark the corresponding circular orbits with the same semi-major axis. The dot-dashed black trajectory marks the fitted geometry of the dust ring, with the fitted eccentricity $e_{\rm ring}$ and semi-major axis $a_{\rm ring}$ labeled in each panel. For ${\rm St}_0 = 10^{-2}$ grains, we additionally calculate the effective alpha viscosity parameters $\alpha_{\rm eff}$ from the radial width of the dust rings and label them in the bottom panels.
\label{fig:xy_sigmad}}
\end{figure*}

While the eccentric response of the gas disk to an eccentric planet is not expected to be as strong as the planet's orbital eccentricity, as shown in Section~\ref{sec:gas_disk}, the eccentric response of the dust disk can be even weaker because dust grains are not necessarily well coupled to the gas. To examine the eccentricity of the dust disk, particularly that of the dust rings at the outer gap edge, we perform a new set of ``dust ring'' simulations, in contrast to the ``full disk'' simulations discussed in the previous sections. Because the dust ring in the full-disk simulations with a Saturn-like planet gradually disappears as the planet eccentricity increases (Figure~\ref{fig:ra_sigmad}), we perform these additional simulations only for the Jupiter-like planet.

\subsubsection{Model setup of the dust-ring simulations}

In our initial attempts, we aimed to use the gas component from the full-disk simulations and initialize a dust ring at the pressure maximum. However, we found that a well-defined pressure maximum is difficult to identify in individual azimuthal slices of the gas disk because of fluctuations in the gas surface density. In addition, the dust distribution does not accurately trace the pressure maximum because of the filtration effect on moderately coupled dust grains (${\rm St}_0 \lesssim 10^{-3}$) described by \citet{bi.J.2024.08}. 

We therefore initialized a new full-disk model following the procedure described in Section~\ref{sec:methods}, but changed the reference Stokes number to ${\rm St}_0 = 10^{-1}$. With this larger Stokes number, dust grains undergo efficient radial drift and will either become trapped at the pressure maximum or leave the computational domain through the inner boundary after 1000 orbits. We then determine the full width at half maximum (FWHM) of the resulting dust ring in each azimuthal slice, denoted by $w_{\rm ring}^{\rm ini}(\phi)$, and define the initial ring radius $R_{\rm ring}^{\rm ini}(\phi)$ as the midpoint between the two half-maximum locations. These two quantities are then used to initialize a new dust ring for the dust-ring simulations. Specifically, in each azimuthal slice, we set the dust density as
\begin{align}
\begin{aligned}
    &\rho_{\rm ring}^{\rm ini}(R,\phi,Z) = \frac{\Sigma_{\rm ring}^{\rm ini}(R_{\rm ring}^{\rm ini}(\phi),\phi)}{\sqrt{2\pi}H_{\rm ring}^{\rm ini}(\phi)}\times\\
    &\exp{\left[-\frac{1}{2}\left(\frac{R-R_{\rm ring}^{\rm ini}(\phi)}{\sigma_{\rm ring}^{\rm ini}(\phi)}\right)^2-\frac{1}{2}\left(\frac{Z}{H_{\rm ring}^{\rm ini}(\phi)}\right)^2\right]},
\end{aligned}
\end{align}
where
\begin{equation} \label{eq:sigma_ring}
    \Sigma_{\rm ring}^{\rm ini}(R_{\rm ring}^{\rm ini}(\phi),\phi) = 0.1\times\Sigma_{\rm g}(R_{\rm ring}^{\rm ini}(\phi),\phi)
\end{equation}
is specified separately for each azimuthal slice,
\begin{equation} \label{eq:H_ring}
    H_{\rm ring}^{\rm ini}(\phi) = 0.2\times h_{\rm g0}\left(R_{\rm ring}^{\rm ini}(\phi)\right)^{9/7}
\end{equation}
is set to 20\% of the local gas scale height at the peak of the dust ring, and
\begin{equation}
    \sigma_{\rm ring}^{\rm ini}(\phi) = \frac{w_{\rm ring}^{\rm ini}(\phi)}{2\sqrt{2\ln2}}
\end{equation}
is chosen to preserve the FWHM of the narrow dust ring formed by the ${\rm St}_0 = 10^{-1}$ grains.

Based on Equations \ref{eq:sigma_ring} and \ref{eq:H_ring}, we do not expect dust back-reaction in the ring to be significant. We therefore divide the total dust mass equally among three dust species with ${\rm St}_0 = 10^{-4}$, $10^{-3}$, and $10^{-2}$ for numerical convenience, instead of adopting a more finely tuned mass allocation based on the MRN distribution \citep{mathis.JS.1977.10}. Because the equilibrium locations and widths of dust rings are also difficult to determine a priori, particularly for smaller grains that are susceptible to filtration, we initialize all three dust species with the same spatial distribution, Keplerian azimuthal velocities, and zero radial and polar velocities. The dust rings are then allowed to evolve toward their corresponding equilibrium states, and we evolve the full model for an additional 1000 orbits.

\subsubsection{Results of the dust-ring simulations} \label{sec:ring_model}

Figure~\ref{fig:xy_sigmad} shows the dust surface density in the dust-ring simulations. To measure the eccentricity and semi-major axis of the dust ring, we first convolve the dust surface density with a Gaussian kernel with an FWHM of $H_{\rm g0}$, in order to smooth out planetary wake patterns that could otherwise bias the ring-fitting procedure. We then define the radial location of the ring in each azimuthal slice as the center of the region corresponding to the highest 50\% of the dust surface density. Finally, we fit these ring locations with an ellipse whose one focus is fixed at the center of mass of the planet--star system.

The fitting results suggest that dust rings composed of grains with Stokes numbers ranging from $10^{-4}$ to $10^{-2}$ all exhibit eccentricities, $e_{\rm ring}$, that agree reasonably well with the coherent eccentricity of the gas disk, $e_{\rm coh}$, at the corresponding semi-major axes, $a_{\rm ring}$, as shown in Figure~\ref{fig:ry_ecccoh}. Consistent with our previous finding in \citet{bi.J.2023.01}, the narrow dust rings initialized at the pressure bump retain a finite radial width, $w_{\rm ring}$, even in the absence of dust diffusion, owing to the effective diffusion induced by planet-disk interactions. Comparing the panels from top to bottom in Figure~\ref{fig:xy_sigmad}, we find that dust rings composed of smaller grains typically have larger semi-major axes, due to the dust filtration effect described in \citet{bi.J.2024.08}. Comparing the panels from left to right, we further find that a higher planet eccentricity tends to increase both $a_{\rm ring}$ and $w_{\rm ring}$.

One additional diagnostic that can be obtained from the dust-ring simulations is the effective alpha-viscosity parameter $\alpha_{\rm eff}$ induced by planet-disk interactions. This parameter regulates $w_{\rm ring}$, relative to a certain characteristic length scale of the radial gas profile $w_{\rm g}$. However, applying this analysis to our models is nontrivial for several reasons. First, previous studies commonly assume a Gaussian radial gas profile and use its FWHM as the width $w_{\rm g}$. In our models, however, the gas profile outside the gap edge is not well described by a Gaussian, as it follows an underlying power-law distribution. Second, individual azimuthal slices of the gas disk are strongly disturbed by planetary wakes and other small-scale structures, making analytic fitting difficult.

To overcome these issues, we first azimuthally average the gas and dust surface density profiles, excluding the azimuthal slices near the planet, to smooth out small-scale structures. We then fit a Gaussian profile to the dust surface density to obtain $w_{\rm ring}$ and $a_{\rm ring}$. The resulting $a_{\rm ring}$ is consistent with that obtained from ellipse fitting to the centers of the half-maximum regions in individual azimuthal slices, as described at the beginning of Section~\ref{sec:ring_model}, with a difference of $\Delta a_{\rm ring} \lesssim 0.01\,R_0$. We use Gaussian fitting here, rather than the ellipse-fitting procedure, to remain consistent with the prescription used for the gas profile below; the ellipse-fitting method is nevertheless required when measuring the ring eccentricity, which cannot be obtained after azimuthal averaging. We next fit a Gaussian profile to the portion of the azimuthally averaged gas surface density profile between $R_0$ and $a_{\rm ring}$ to obtain $w_{\rm g}$. Finally, using ${\rm St}(a_{\rm ring})$, the midplane Stokes number at the semi-major axis of the dust ring, we estimate $\alpha_{\rm eff}$ using Equation 46 in \citet{dullemond.CP.2018.12}:
\begin{equation}
    \alpha_{\rm eff} = {\rm St}(a_{\rm ring})\times\left(\frac{w_{\rm ring}}{w_{\rm g}}\right)^2.
\end{equation}

Moreover, because only the ${\rm St}_0 = 10^{-2}$ grains reach a quasi-steady equilibrium within the simulation time of a few thousand orbits, we perform this analysis only for the largest grains, shown in the bottom panels of Figure~\ref{fig:xy_sigmad}. The results suggest that a Jupiter-like planet on a moderately eccentric orbit can induce a relatively high effective viscosity, $\alpha_{\rm eff}\sim 10^{-3}$, which increases with planet eccentricity. The corresponding values of $\alpha_{\rm eff}/{\rm St}$ are 0.06, 0.12, 0.25, and 0.44 for $e_{\rm p} = 0$, 0.05, 0.1, and 0.2, respectively. These values are comparable to those inferred from dust-trapping models applied to ALMA observations of protoplanetary disks (e.g., Figure~7 in \citealt{dullemond.CP.2018.12}). However, we note that all these estimated values should be interpreted with caution, because they are obtained within a limited parameter space and rely on several assumptions and approximations that may not capture the full complexity of real disks.

\subsection{Insights into disk observations}

Recently, \citet{close.LM.2025.08} reported the detection of a $\sim$$5\,M_{\rm J}$ planet, WISPIT~2b, in the WISPIT~2 system at a deprojected distance of $\sim$54 au from the star. The planet was detected through H$\alpha$ emission imaged with the MagAO-X instrument on the Magellan Clay Telescope. Meanwhile, \citet{vancapelleveen.RF.2025.08} reported that the system hosts a protoplanetary disk observed in near-infrared (NIR) scattered light with the SPHERE instrument on the Very Large Telescope. The disk contains a gap co-located with the planet, and at least two dust rings exterior to the planet's orbit, at $\sim$100 and $\sim$160 au, respectively.

Orbit modeling by \citet{vancapelleveen.RF.2025.08} suggests that the planet most likely has a semi-major axis of $a_{\rm p}$ = 57 au and a low eccentricity of $e_{\rm p} \lesssim 0.2$. This configuration may explain the formation of the co-located NIR gap and the NIR ring at $\sim$1.75$\,a_{\rm p}$, but it does not provide a clear explanation for the second ring at $\sim$2.8$\,a_{\rm p}$. The situation is further complicated by the follow-up ALMA observations reported by \citet{facchini.S.2026.02}. These observations, obtained at millimeter wavelengths with a spatial resolution of 2 au, reveal only an extremely narrow dust ring with $\sigma_{\rm ring}$ = 7.2 au at $\sim$144 au ($\sim$2.5$\,a_{\rm p}$), and show no corresponding emission near $\sim$100 au. Possible explanations proposed by \citet{facchini.S.2026.02} include an additional planet at $\sim$130 au, a substantially higher mass of $15\,M_{\rm J}$ for WISPIT~2b than previously estimated, or a moderate orbital eccentricity.

While our models do not directly address the NIR observations because they neglect dust diffusion that sustains dust layers at the disk surface beyond the planet-opened gap, they can still provide useful insight into the dust ring observed by ALMA. We also note that the estimated mass of WISPIT~2b lies beyond the parameter range explored in our simulations. From the perspective of our models, the extremely narrow ALMA dust ring at $\sim$144 au appears inconsistent with a planet more massive than Jupiter on a moderately eccentric orbit. Even if we assume that the ALMA dust ring is located at $\sim$2$\,a_{\rm p}$, which is already farther out than the ring locations in Figure~\ref{fig:xy_sigmad}, and further assume optimistically that the planet is observed at periapsis, the planet eccentricity would still need to be $e_{\rm p} \gtrsim 0.1$. However, such a planet would be expected to produce a much wider dust ring than observed. The ring would also likely exhibit a noticeable eccentricity in such cases, which has not been reported in either the NIR or ALMA observations.

Considering that a gap is observed at $\sim$130 au in NIR observations, we tentatively suggest that an additional planet at this location may provide a more plausible explanation. Testing this possibility will require new simulations with multiple planets and dust diffusion to better model the dust distribution at the disk surface and thereby provide more direct insight into the NIR observations. More generally, we will also investigate the role of planet multiplicity in shaping dust puff-up features in the third paper of this series. Such a study will be relevant not only to the WISPIT~2 system, where a second planet has already been confirmed at a much smaller orbital radius of $\sim$14 au \citep{lawlor.C.2026.04}, but also to other well-studied systems such as PDS~70, where two planets have been confirmed, and an ALMA dust ring with an unusual radial profile, consisting of a ring plus a shoulder, has been observed \citep{keppler.M.2018.09,haffert.SY.2019.06,benisty.M.2021.07}.

\subsection{Insights into planetesimal formation}

Dust rings in the disk are often considered favorable sites for planetesimal formation, because the enhanced dust-to-gas ratio in these rings facilitates more efficient grain growth through coagulation and further promotes the conditions required for triggering the streaming instability (SI) and subsequent planetesimal formation \citep{youdin.AN.2005.02,johansen.A.2007.06,lim.J.2025.11,lim.J.2026.04}. However, our study raises the question of whether dust rings formed by eccentric planets remain similarly favorable sites for planetesimal formation.

On the one hand, as shown in Figure~\ref{fig:xy_sigmad}, dust rings formed by eccentric planets tend to experience stronger turbulence, as parameterized by $\alpha_{\rm eff}$. This may be unfavorable for planetesimal formation because stronger turbulence can increase the collision velocities between dust grains, promoting fragmentation over coagulation \citep{ormel.CW.2007.05,jiang.H.2024.02}. In addition, several previous studies have suggested that SI can be suppressed in turbulent environments \citep{chen.K.2020.03,lim.J.2024.07}. On the other hand, Figures~\ref{fig:rz_rhod2g}, \ref{fig:ra_sigmad}, and \ref{fig:xy_sigmad} show that dust rings in the eccentric-planet cases tend to have lower dust-to-gas ratios and dust surface densities than those in the circular-planet cases, which can further hinder grain growth and the growth of SI. Therefore, dust rings formed by eccentric planets may be less favorable sites for planetesimal formation than those formed by circular planets. And how much less favorable they are will depend on the degree of planet eccentricity, which is to be explored in future studies. 

\section{Conclusions} \label{sec:conclude}

In this paper, we investigated the role of a planet's orbital eccentricity in shaping gas and dust structures in protoplanetary disks through planet-disk interactions using three-dimensional multifluid hydrodynamic simulations. We found that

\begin{enumerate}
    \item Consistent with previous 2D studies, eccentric planets can drive eccentric motion in the gas disk, but the eccentricity of the gas disk is generally smaller than that of the planet. Although the gap opened by an eccentric planet also becomes eccentric, azimuthal averaging remains a good approximation for measuring the gap depth and location, because the gap eccentricity remains modest in our simulations.
    \item The dust puff-up feature can be amplified by the planet's orbital eccentricity. This occurs because the eccentric planet drives eccentric motion in the gas disk, which produces stronger meridional gas circulation than in the circular-orbit case and consequently lofts dust grains to higher disk elevations.
    \item The gap opened by an eccentric planet can become highly leaky to dust grains, and the degree of leakage increases with planet eccentricity. This occurs because the radial extent of the horseshoe streamlines becomes large enough for an eccentric planet that the streamlines connect the gap with the surrounding dust disk, allowing dust grains to be radially transported across the gap.
    \item Dust rings composed of pebble-sized grains are expected to become both larger and radially wider when the planet is eccentric, with this trend becoming more pronounced at higher planet eccentricities. This poses a challenge for explaining the extremely narrow dust ring observed in the WISPIT~2 disk with only one massive planet on a moderately eccentric orbit.
\end{enumerate}

We thank the anonymous referee for insightful comments, and Mario Flock, Douglas Lin, and Zhaohuan Zhu for helpful discussions, which greatly improved the quality of this study. The numerical simulations presented in this work were carried out on the KAWAS cluster hosted by the Academia Sinica Institute of Astronomy and Astrophysics. J.B. is supported by the Deutsche Forschungsgemeinschaft (German Research Foundation, DFG) under grant number 544937803. J.B. further acknowledges funding from the ASIAA Distinguished Postdoctoral Fellowship and COST Action CA22133 PLANETS, supported by the European Cooperation in Science and Technology (COST). M.-K.L. is supported by the National Science and Technology Council (grants 113-2112-M-001-036-, 114-2112-M-001-018-, 113-2124-M-002-003-, and 114-2124-M-002-003-), an Academia Sinica Career Development Award (AS-CDA-110-M06), and an Academia Sinica Grand Challenge Seed Program (AS-GCS-115-M02). M.-K.L. thanks the Tsung-Dao Lee Institute for its hospitality, where part of this work was completed. 

\bibliography{refs}

\begin{appendix}
\section{Morphology of dust puff-up in individual azimuthal slices} \label{sec:d2gazi}

\begin{figure*}
\centering
\includegraphics[width = \textwidth]{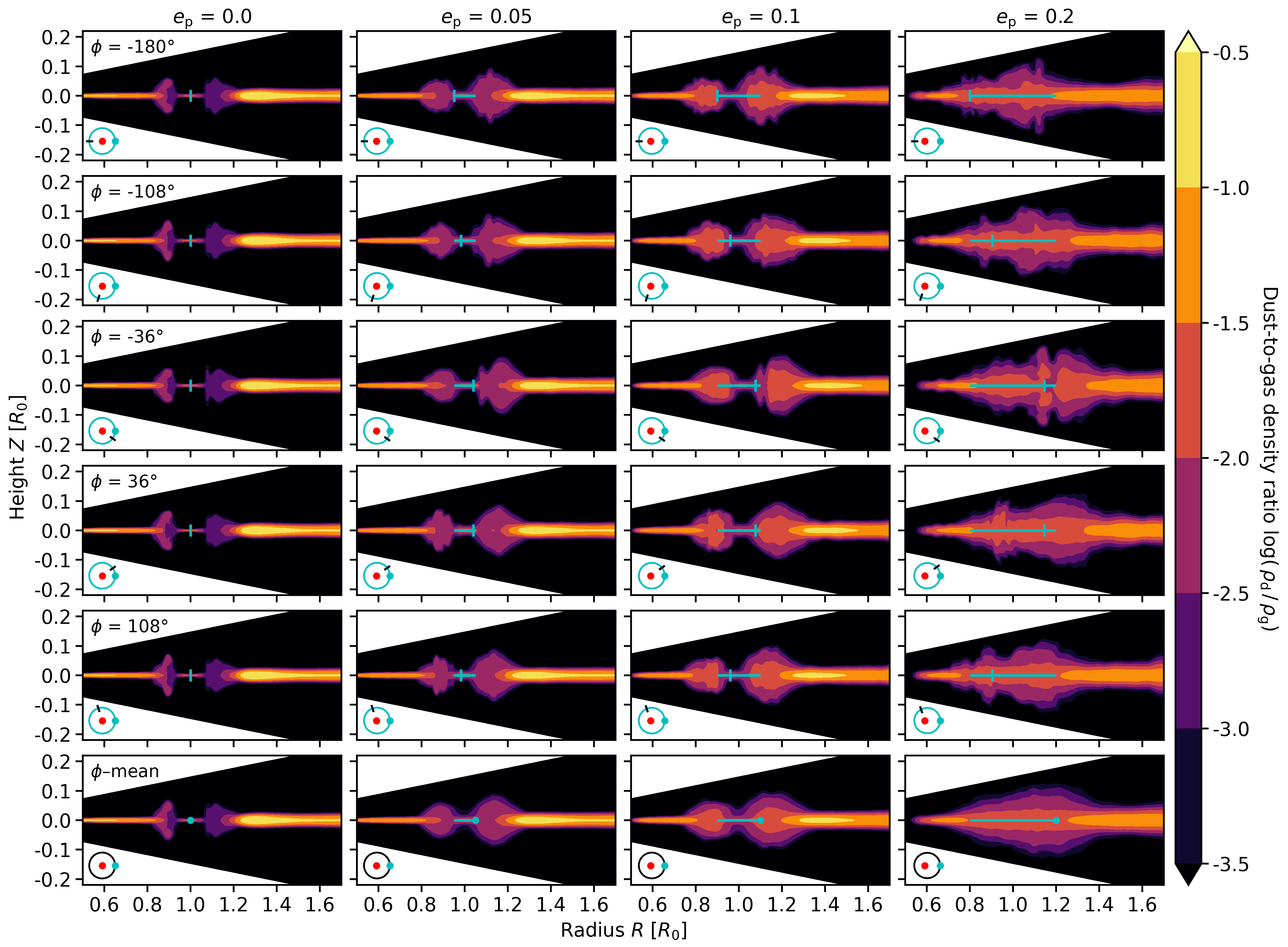}
\figcaption{Dust-to-gas volumetric density ratio at $t = 1000\,T_0$ in models with a Saturn-like planet $M_{\rm p} = 3\times10^{-4}M_\star$ and different orbital eccentricities $e_{\rm p}$, shown at different azimuthal angles $\phi$ measured from the planet's apoapsis in the direction of orbital motion. Panels in the same column share the same $e_{\rm p}$, while panels in the same row share the same $\phi$, except for the bottom row, which shows the azimuthally averaged field. The vertical cyan line in the midplane marks the location of the planet's orbit at the given $\phi$, while the horizontal cyan line marks the radial range traversed by the planet over its orbit. The planet is marked by a cyan dot in the midplane in the bottom panels. A not-to-scale sketch of the planet's orbit is shown in the bottom left corner of each panel: the orbit is marked by a cyan trajectory, the star by a red dot, the planet by a cyan dot, and the azimuthal slice corresponding to each panel by a black line. In the bottom panels, the trajectory is colored black to indicate azimuthal averaging over the full orbit.
\label{fig:rz_d2gsat}}
\end{figure*}

\begin{figure*}
\centering
\includegraphics[width = \textwidth]{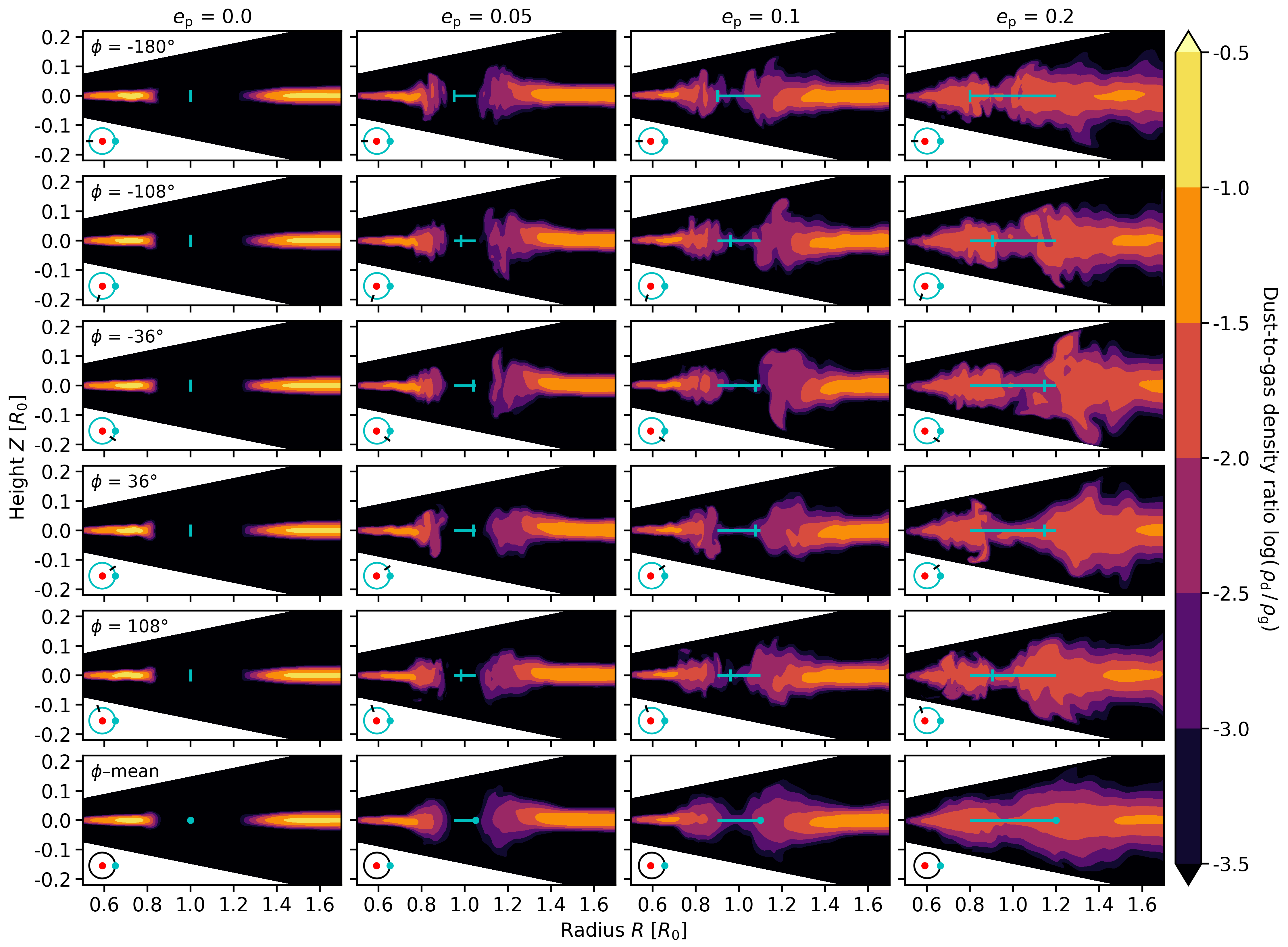}
\figcaption{Similar to Figure~\ref{fig:rz_d2gsat}, but in models with a Jupiter-like planet $M_{\rm p} = 10^{-3}M_\star$. 
\label{fig:rz_d2gjup}}
\end{figure*}

Figure~\ref{fig:rz_d2gsat} and Figure~\ref{fig:rz_d2gjup} show the dust-to-gas volumetric density ratio in the $R$--$Z$ plane at different azimuthal angles, $\phi$, for models with a Saturn-like and a Jupiter-like planet, respectively. The azimuthal domain is divided into five equal intervals in $\phi$, and the slices are sampled starting from the periapsis of the planet's orbit to avoid the apoapsis slice, where the planet is located. Both figures show that the dust puff-up feature does not vary significantly with $\phi$, and that the individual azimuthal slices agree reasonably well with the azimuthally averaged fields shown in the bottom panels. This suggests that azimuthal averaging remains a good approximation for this study.

\end{appendix}

\end{document}